# Van der Waals injection-molded crystals

Vinh Tran[1†], Amy X. Wu[1†], Laisi Chen[1], Ziyu Feng[1], Vijay Kumar[1], Takashi Taniguichi[2], Kenji Watanabe[2], Javier Sanchez-Yamagishi[1*]

1. Department of Physics and Astronomy, University of California Irvine, Irvine, CA, USA
2. Research Center for Electronic and Optical Materials, National Institute for Materials Science, 1-1 Namiki, Tsukuba 305-0044, Japan

†these authors contributed equally to this work
*corresponding author
Email: javier.sanchezyamagishi@uci.edu

## Abstract

Shaping low-dimensional crystals into precise geometries with low disorder is an outstanding challenge. Here, we present a method to grow single crystals of arbitrary geometry within van der Waals (vdW) materials. By injecting molten material between atomically-flat vdW layers within an $SiO_2$ mold, we produce ultraflat and thin crystals of bismuth, tin, and indium that are shaped as hallbars, rings, and nanowires. The crystals are grown fully encapsulated in hexagonal boron nitride, a vdW material, providing protection from oxidation. Varying the depth of the mold allows us to control the crystal thickness from ten to a hundred nanometers. Structural measurements demonstrate large single-crystals encompassing the entire mold geometry, while transport measurements show reduced disorder scattering. This approach offers a means to produce complex single-crystal nanostructures without the disorder introduced by post-growth nanofabrication.

## Main Text

Confining a material into low dimensions strongly modifies its electronic structure by opening band gaps[1], reducing the number of bulk modes compared to surfaces or edges[2,3], and inducing topological phase transitions. These effects are key to realizing engineered quantum systems such as quantum dots or Majorana bound states[4]. The challenge in realizing low-dimensional materials is retaining crystallinity and uniformity while reducing material dimensions. For a narrow range of layered materials, such as graphite and transition metal dichalcogenides, achieving two-dimensional crystals is straightforward[5]. However, these approaches are inapplicable to a vast majority of materials that prefer three-dimensional structures. As a result, it is challenging to produce uniform crystals of nanometer scale thickness, where conventional deposition techniques such as molecular beam epitaxy or chemical vapor deposition result in thin films with irregular surfaces, small domain structure, or undesirable surface interactions[6–9].

Lateral confinement into one or zero-dimensional structures introduces further challenges. Top-down approaches such as reactive ion etching or ion beam milling introduce disorder and contamination[10,11]. Bottom up synthesis methods such as vapor-liquid-solid[12,13] or epitaxial growth[14] can avoid these contamination issues, but with the drawback of reduced uniformity and limited control of morphology.

An alternative approach is to directly grow crystals into a confined geometry via injection molding. Here, heat and pressure are applied to drive source material into a confined mold. In principle, this approach can create low-dimensional crystals with defined geometry without the need for post-fabrication processing. Approaches based on this principle have been used to grow nanowires of metals[15–18] and topological materials[19,20], as well as quasi-2D flakes[21] using aluminum oxide or silicon as the molding surface. However, this approach has been limited by the irregularities in the mold surfaces, challenges in removing the crystals from the mold and protecting the crystals post-synthesis.

Van der Waals (vdW) materials are an appealing choice as a molding substrate, as they naturally cleave into atomically smooth surfaces free of dangling bonds[22]. Moreover, vdW materials form hermetic interfaces that can protect air-sensitive materials[23,24]. In particular, hexagonal Boron Nitride (hBN), a wide bandgap vdW material, has demonstrated extensive use as an encapsulant[25,26]. Recently, our group demonstrated the confined growth of thin bismuth crystals in between layers of hBN via melt growth under compression[27]. The resulting samples had large single crystal domains and exhibited enhanced transport properties as compared to MBE-grown thin films. Other recent works have also demonstrated the growth of ultrathin or even two-dimensional crystals between vdW materials[28–30]. However, these previous demonstrations have had limited control of thickness or lateral crystal geometry.

Here, we demonstrate a technique to grow thin single crystals of predefined geometry by injecting molten material into a vdW-lined mold, a process we refer to as vdW injection molding. Our approach enables the direct production of ultraflat single crystals of bismuth, tin, and indium in various geometries, such as hall bars and nanowires, which are fully encapsulated within a vdW material. The combination of geometric control, quantum confinement, low disorder, and protection from oxidation unlocks a new regime for single-crystal quantum devices.

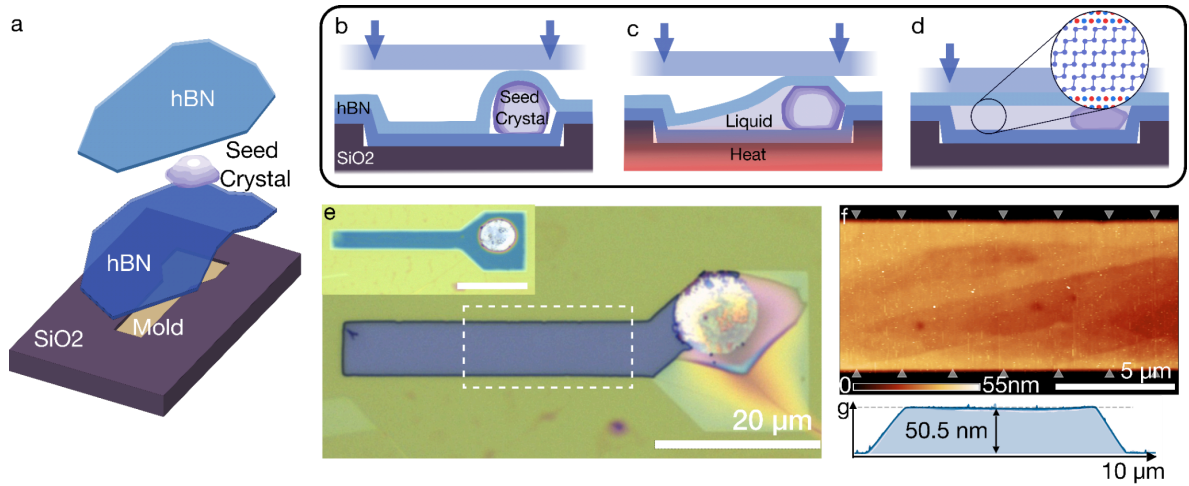

***Fig 1. Synthesizing low-dimensional crystals by van der Waals injection molding.***
*(a) Schematic of a vdW injection mold consisting of an hBN encapsulated source flake on top of an etched SiO$_2$ trench. (b-d) Schematics showing the injection molding process: (b) initial state, (c) melting under compression, and (d) cooling down while maintaining pressure. (e) Optical image of a 55 nm thick bismuth vdW injection-molded sample (S57). Inset shows an optical image of the completed vdW injection mold before the squeezing step. (f) AFM topography of the crystal surface showing large terraces through the top hBN. Measured region is indicated by the white box in (e). (g) Height profiles measured across the sample width indicated by the white arrows in (f).*

      To create crystals by vdW injection molding, we force molten material into a predefined mold that is lined with hBN (Fig 1a). The injection mold consists of a trench etched into thermally-grown SiO$_2$ on a Si substrate. We then transfer into the trench a flake of source material encapsulated in hBN (additional details in the supplementary info - methods section). After transferring, the hBN layers conform to the shape of the trenches (Fig 1e, inset). Injection molding is then achieved by melting the source material while compressing it into the mold with a flat top substrate (Fig 1b-d). The stack is then slowly cooled to crystallize under compression.

      Injection molding directs the molten material away from its initial oxide shell, resulting in more uniform crystals than the undirected molding we achieved previously[27]. A typical result is shown in Fig 1e, where the bismuth that began at the far right of the trench now fills the entire rectangular mold, resulting in a crystal of uniform color. On the right side of the trench, at the location of the original bismuth flake, a discolored oxide shell remains. We find that the oxide shell does not move significantly from its initial position during the injection molding process. Thus, this method allows us to grow clean crystals in a confined environment while minimizing the presence of oxide slag. Furthermore, the encapsulation in hBN protects the crystal from additional oxidation, as evidenced by the absence of additional oxide formation if the encapsulated crystals are re-melted (SI Fig S12).

      Atomic Force Microscopy (AFM) measurements of the sample show a 55 nm tall mesa-like height profile that has less than a 3% height variation across the 10 μm x 7 μm channel (Fig 1f-g). The surface of the bismuth crystal features large terraces several microns wide, resolvable through the top hBN layers.

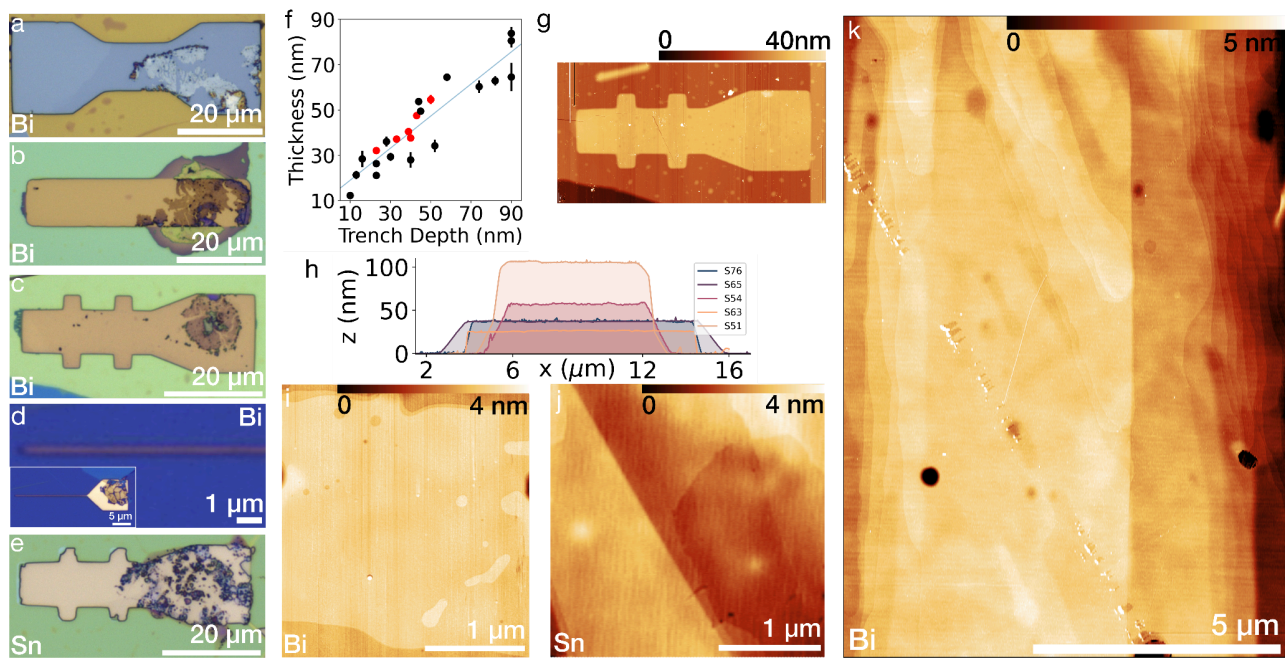

***Fig 2. Control of crystal geometry by vdW injection molding***
*(a-e) Optical images of various injection molded crystals with different shapes showing well defined lateral geometry (samples S29, S42, S76, LV18, and Sn13, respectively). (f) Sample thickness plotted against trench depth for bismuth samples with a linear fit shown (blue line). Bar length represents standard deviation from the mean. Samples highlighted in red are molded with a top sapphire substrate. (g) AFM topography image of a hallbar crystal (S76). (h) Line cuts across the width of five samples. (i-k) AFM topography of crystal surfaces after removing top hBN showing large terraces (samples S55, Sn02, and S58, respectively).*

      VdW injection molding allows us to synthesize crystals of bismuth, tin and indium with a wide range of planar geometries. To highlight the in-plane geometry control, we fabricated 25 samples of different geometries ranging from 150 nm to 30 μm in width and up to 70 μm in length. The geometries include bowties, rectangles, hallbars, nanowires, and rings as shown in Fig 2a-e and Fig S2-S5.

      For bismuth, the etched $SiO_2$ mold provides excellent control over the lateral dimensions of the crystal. The dimensions of the molded crystals match those of the original $SiO_2$ mold within the optical resolution of our microscope (Fig S4). Under compression, the mold traps the injected material until it is fully filled, at which point excess material tends to be ejected out of the mold through the hBN layers (SI Video 2).

      For indium and tin samples, we can also achieve lateral geometry control (Fig 2e and Fig S5). However, the resulting crystals tend to have faceted irregularities along the edges and from internal voids. These irregularities can change over time, especially in thinner crystals and under the effects of heating. We speculate that the encapsulated metal crystals are free of surface oxides, making them prone to deformation creep under the confining pressure of the hBN. Such a process may be more prevalent in tin and indium compared to bismuth due to

higher surface energies and lower melting points[31,32]. Further discussion can be found in the caption for Fig S5.

This method provides control over sample thickness, as we find that the depth of the etched $SiO_2$ trench determines the thickness of the molded crystals. By varying the depths of the trench, we produce crystals of thicknesses from 11 to 106 nm, with an approximately linear relationship between crystal thickness and trench depth (Fig 2f). This thickness-trench depth correlation is independent of the thickness of the hBN used.

Fig 2h shows the flat mesa-like profile of five uniform crystals, with thicknesses ranging from 26 to 106 nm thick. For these samples, the largest variation is along the width of the crystal, with the middle of the crystal being on average 1.65% taller than the sides. This corresponds to a slope of 0.2 ± 0.14 nm/µm along the width, i.e. less than one atomic layer per micron. For these uniform crystals, we find a similar average slope along the channel length of 0.14 ± 0.17 nm/µm. Overall, the trench-molded crystals are significantly flatter than crystals grown on a flat surface without a mold, which tend to result in dome-shaped crystals[27].

The uniformity of the injection-molded crystals is sensitive to the squeezing pressure profile and rigidity of the compression substrates. By using top-squeezing substrates made of sapphire, we produce bismuth crystals with a thickness that is uniform within 1.7 to 3.6% (Fig 2f, red points). When compressing with glass top substrates, we observe substantially more non-uniform crystals with either concave or convex profiles along the crystal width (Fig S7). We ascribe this to the lower Young's modulus of glass (80 GPa) compared to sapphire (400 GPa), resulting in deformation of the squeezing substrate during injection molding. We anticipate that further improvements in crystal uniformity would be achieved by also replacing the bottom substrate with sapphire as well as by using thinner hBN crystals due to the low 36 GPa Young's modulus of hBN[33].

The advantage of vdW molding is that the atomically flat surface of the vdW material is imprinted onto the crystal surface. To directly characterize the crystal surface, we mechanically remove the top encapsulating hBN using a polyvinyl chloride-covered stamp[34]. AFM of the Bi and Sn surfaces reveals flat terraces 1 to 10 um in width (Fig 2i-k). Within the single terraces of Bi and Sn (Fig 2i and 2j), we measure root mean squared roughness values of 0.11 nm and 0.09 nm, respectively (see Fig S8), comparable to the 0.05 nm roughness of hBN itself [27]. The large terraces tend to be in the center of the crystals, with narrow and denser terraces at the edges of the crystal. Occasionally present are some small, sparsely spaced holes that tend to be 500 nm or smaller in diameter with depths less than the total crystal thickness. We speculate these holes originate from surface imperfections in the substrates or trapped gases.

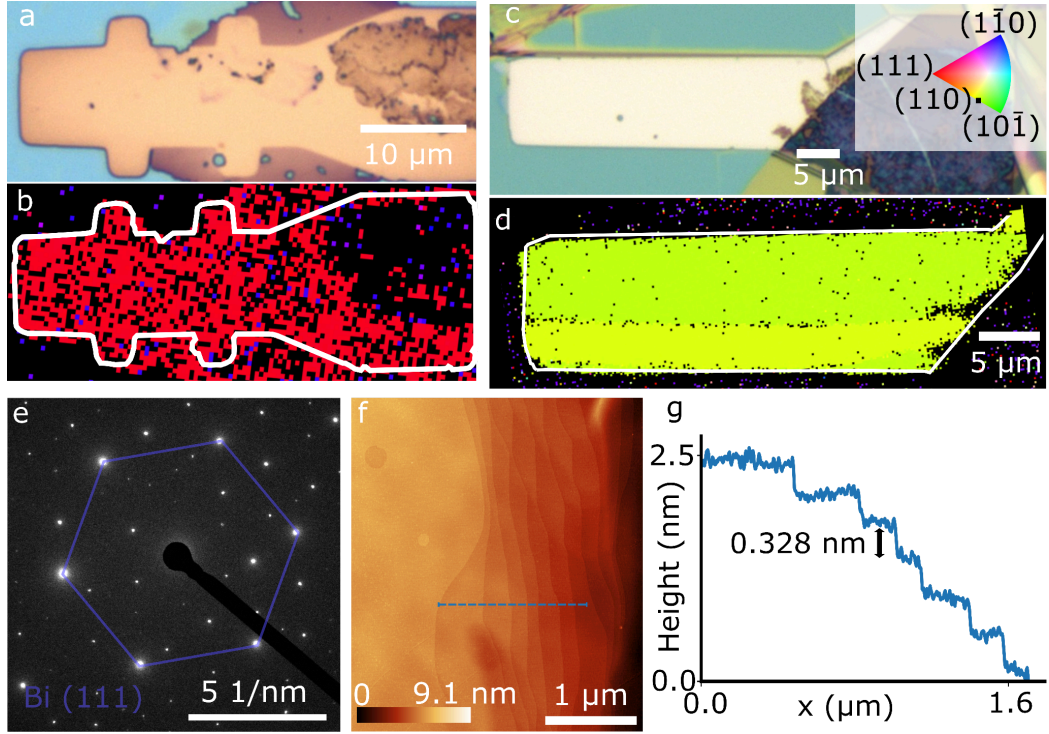

***Fig 3. Crystallinity of vdW injection molded crystals. (a)** Optical image of injection-molded bismuth sample S79 and **(b)** EBSD inverse pole figure (IPF Z) of the same sample showing that it is single crystal and in the (111) orientation. **(c)** Optical image of sample S58 after removing its top hBN flake and **(d)** IPF Z of the same sample showing two large domains with a (110) orientation. **(e)** TEM selected area diffraction measurement of injection-molded bismuth (S68) after its top hBN has been removed showing a (111) diffraction pattern. **(f)** AFM topography of the injection-molded bismuth (S58) after the top hBN flake has been removed showing step terraces. **(g)** Height line trace taken from (f). Average step height of 0.33 +/- 0.05 nm matches the expected step height of 0.328 nm for buckled (110)[35] bilayers. Step height measured by averaging 31 line traces from (f) (SI Fig S9).*

Diffraction measurements show that vdW injection-molding results in large single-crystal domains that can comprise the entire mold geometry (Fig 3a-b). All of the samples we measured exhibit a uniform out-of-plane crystal orientation throughout the mold. 7 out of 8 crystals were single crystals, as verified by EBSD or TEM diffraction measurements showing uniform in-plane crystal orientation that is unchanged across the sample (see Fig 3e, S10-S11). Only one of the crystals had an internal domain wall, which runs parallel to the crystal length and divides two domains of opposite in-plane orientation (Fig 3d).

Diffraction and AFM measurements show that vdW injection molding results in bismuth single crystals of both hexagonal (111) and square-like (110) thin film orientations (miller indices refer to the rhombohedral unit cell). These surface orientations feature different crystal symmetries and electronic structures, but are similar in both having Rashba surfaces states that transition to a 2D topological insulator phase in the 2D limit[36–39]. We found four of our crystals to be (110) and four to be (111) oriented, with no correlation between crystal orientation and thickness. Previously, when molding bismuth with hBN on a flat substrate, we only observed the (111) orientation[27]. By contrast, a recent study showed that bismuth molded between flat $MoS_2$ on sapphire substrates is (110) orientated below 50 nm in thickness[28]. We speculate that substrate interactions, pressure, and the lateral confinement of the trench mold may determine the orientation of the vdW injection-molded crystals.

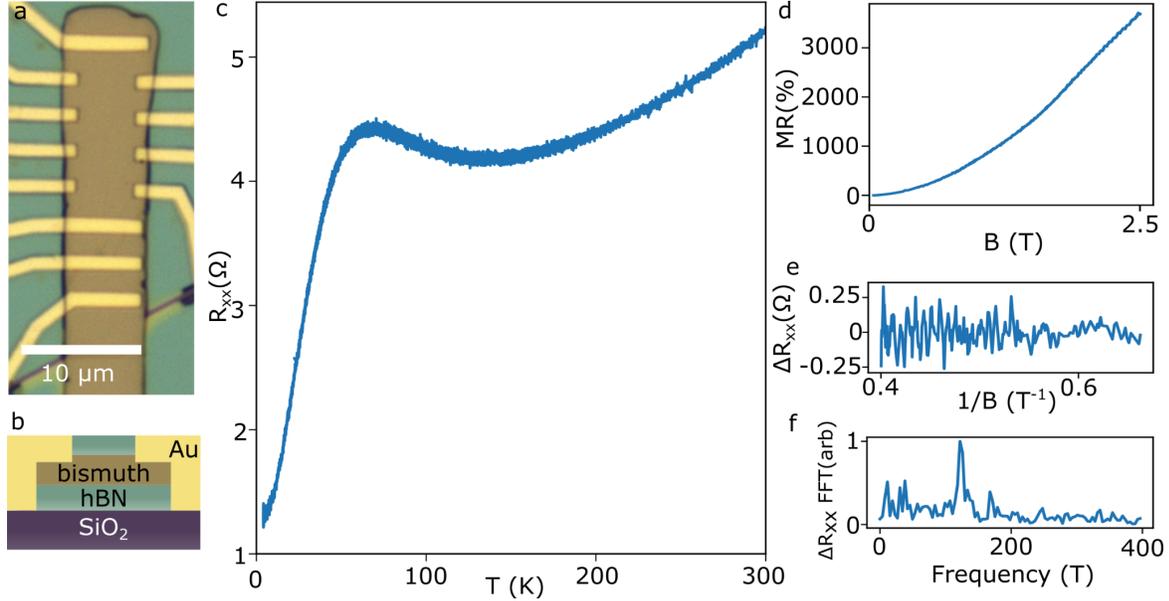

*Fig 4 Transport measurements in vdW injection molded bismuth. (a) Optical image of an injection-molded bismuth device (S51) with Cr/Au electrodes. (b) Schematic showing the side-view cross-section of the device. (c) Resistance vs temperature of the device from 300 K to 1.5 K measured on the electrodes indicated by the inset diagram (the voltage probe distance is 3 μm, channel width is 6 μm, and sample thickness is 106 nm). (d) Magnetoresistance ratio vs out of plane magnetic field measured at 1.5 K. (e) Background subtracted magnetoresistance vs 1/B for magnetic fields from 1.5 T to 2.5 T (top). (f) The fast Fourier transform (FFT) of the resistance oscillations shown in (e).*

The injection-molded bismuth shows high-quality transport properties. In a 106 nm thick sample, we observe a 4x residual resistance ratio (RRR), which is larger than is achieved in MBE-grown thin films[40,41] and is due to the low residual resistivity at 4 K of 6.94e-6 Ω-cm. The overall temperature dependence is metallic, which we attribute to the bismuth surface states and Bi bulk bands that are not fully gapped out for this relatively thick sample. Consistent with the improved RRR, we observe a large magnetoresistance of 3686% at a field of 2.5 T. Resistance quantum oscillations onset at a field 2 T, which is lower than the 3 T onset field that we measured previously in vdW-molded bismuth[27]. Together with the improved RRR, this suggests lower disorder scattering as compared to previous thin film studies. We conclude that molding enables the measurement of high-quality crystals of controlled geometry without the use of post-growth etching techniques.

In summary, we have demonstrated a method to grow hBN-encapsulated crystals of various geometries by vdW injection molding. Growing between hBN vdW layers simultaneously enables ultraflat crystal surfaces, protection from oxidation via encapsulation, facile transfer onto arbitrary substrates, and enhanced electronic transport - resolving many of the previous limitations of molded crystal growth. These features, previously unique to vdW materials, can be adopted by generic materials through vdW injection molding.

The ability to grow low-disorder nanocrystals of arbitrary shape while protected from oxidation has the potential to open new frontiers in the study of topological materials and clean heterointerfaces with superconductors. Of the materials studied so far, both bismuth and alpha-tin support various topological phases, while indium and beta-phase tin are superconductors. Due to the thermal and chemical stability of hBN, this technique should be generalizable to higher melting point materials beyond monoelemental compounds. We envision that vdW injection molding and related techniques can enable complex crystalline structures with far lower disorder than conventional nanofabrication techniques, with potential applications in metamaterials, photonics, and quantum computing.

## Supporting Information

Additional experimental details, materials, methods (pdf)
Supporting Information Video 1 - injection molding process (mp4)
Supporting Information Video 2 - injection molding process (mp4)

# Author Information

## Corresponding Author

Corresponding Author: Javier D. Sanchez-Yamagishi
Email: javier.sanchezyamagishi@uci.edu

## Author contributions

J.D.S.-Y. supervised the overall research. V.T., A.X.W., Z.F. and V.K. prepared and characterized the samples. V.T. and A.X.W. performed and analysed electronic measurements. K.W. and T.T. synthesized the hBN samples. V.T., A.X.W., L.C., Z.F., V.K., and J.D.S.-Y. wrote the manuscript.

## Competing Interests

The authors declare no competing financial interest.

# Acknowledgements

We acknowledge the use of facilities and instrumentation at the Integrated Nanosystems Research Facility (INRF), in the Samueli School of Engineering at the University of California

Irvine, and at the UC Irvine Materials Research Institute (IMRI), which is supported in part by the National Science Foundation through the UC Irvine Materials Research Science and Engineering Center (DMR-2011967). We thank Ilya Krivorotov and Xinyao Pei for the assistance and use of their sputtering machine. We thank Mingjie Xu and Toshihiro Aoki for TEM assistance, and Qiyin Lin for electron-beam evaporation assistance. We thank Michaela Bacani and Ian Sequeira for assistance in nanofabrication.

# Funding

The fabrication and measurement of injection molded bismuth devices was supported by the Air Force Office of Scientific Research FA9550-23-1-0454. K.W. and T.T. acknowledge support from the JSPS KAKENHI (Grant Numbers 21H05233 and 23H02052) and World Premier International Research Center Initiative (WPI), MEXT, Japan. V.T. acknowledges funding from a GAANN fellowship from the U.S. Department of Education under contract P200A240014.

# Data Availability

All of the data that support the findings of this study will be published to Zenodo.

# Supporting Information: Van der Waals injection-molded crystals

Vinh Tran[1†], Amy X. Wu[1†], Laisi Chen[1], Ziyu Feng[1], Vijay Kumar[1], Takashi Taniguichi[2], Kenji Watanabe[2], Javier D. Sanchez-Yamagishi[1*]

1. Department of Physics and Astronomy, University of California Irvine, Irvine, CA, USA
2. Research Center for Electronic and Optical Materials, National Institute for Materials Science, Tsukuba, Japan

†these authors contributed equally to this work
*corresponding author
Email: javier.sanchezyamagishi@uci.edu

## Materials and Methods

**Material Source Preparation**

We prepare flakes of source material by compressing grains of bulk material to form 100-500 nm thick discs. For bismuth, powder (Goodfellow, 150-µm-diameter, 99.999% suspended in isopropyl alcohol) is first drop-casted onto an $SiO_2$/Si chip. The chip is then covered with a glass coverslip (VWR Micro Coverglass #2). Pressure is then applied to flatten the grains of powder. Tin (US Research Nanomaterials, 60-80 nm, 99.9%) is prepared similarly. For indium, foil (Sigma Aldrich, 0.1mm thick, 99.995%) is cut into 100 micron to 1 mm sized sheets and transferred onto $SiO_2$/Si chips before being compressed under heat.

**Fabricating Etched $SiO_2$ Trenches**

We create the $SiO_2$ trenches that form the base of the injection mold by selectively etching areas of $SiO_2$/Si chips through a polymethyl methacrylate (PMMA) mask. To create the PMMA mask, we start by spinning a layer of PMMA (Kiyaku 950 A5, 500 nm, 2000 RPM, 2 minutes) onto 90 nm and 300 nm oxide $SiO_2$/Si chips (University Wafer) before baking for 10 minutes at 180 °C. We then perform electron-beam lithography (FEI Magellan 400 XHR SEM) and develop the chips (isopropyl alcohol:Water in ratio of 3:1) to expose regions of $SiO_2$ in the lateral geometry of our molds (Fig. S1a,e).

The exposed $SiO_2$ regions are then etched with a thin layer of a chemical etchant cream (ArmorEtch) (Fig. S1b,f). The etch rate is around 5 nm/s[1]. We tune etch times to achieve desired etch depths (10-90 nm). The chips are then cleaned by sonication in deionized water, acetone, and isopropyl alcohol to remove the PMMA mask and any remaining etchant residue. We then measure the exact etch depth and identify clean molds using atomic force microscopy topography imaging (Park NX10) (Fig. S1i-j). The thermally grown $SiO_2$ is etched quite uniformly (Fig. S1j).

## Transferring hBN Encapsulated Source Flakes

With an etched $SiO_2$ trench completed, we then transfer on a hBN encapsulated source flake using a conventional polycarbonate (PC)/Polydimethylsiloxane (PDMS) transfer technique[2]. First, bulk hBN is mechanically exfoliated onto 300 nm oxide $SiO_2$/Si wafers and 5-60 nm flakes with areas larger than the trench area are selected using optical microscopy. Then the individual layers of the hBN/source flake/hBN stack are individually picked up using a PC film on a PDMS stamp and transferred onto the etched $SiO_2$ mold, building the encapsulated structure up layer by layer. After each step, the PC film is dissolved in chloroform (Fig. S1c,d,g,h). The hBN layers are sufficiently thin such that they collapse into and conform to the mold, acting as a liner (Fig S1k-l). This completes the fabrication of the vdW injection mold structure (Fig S1h).

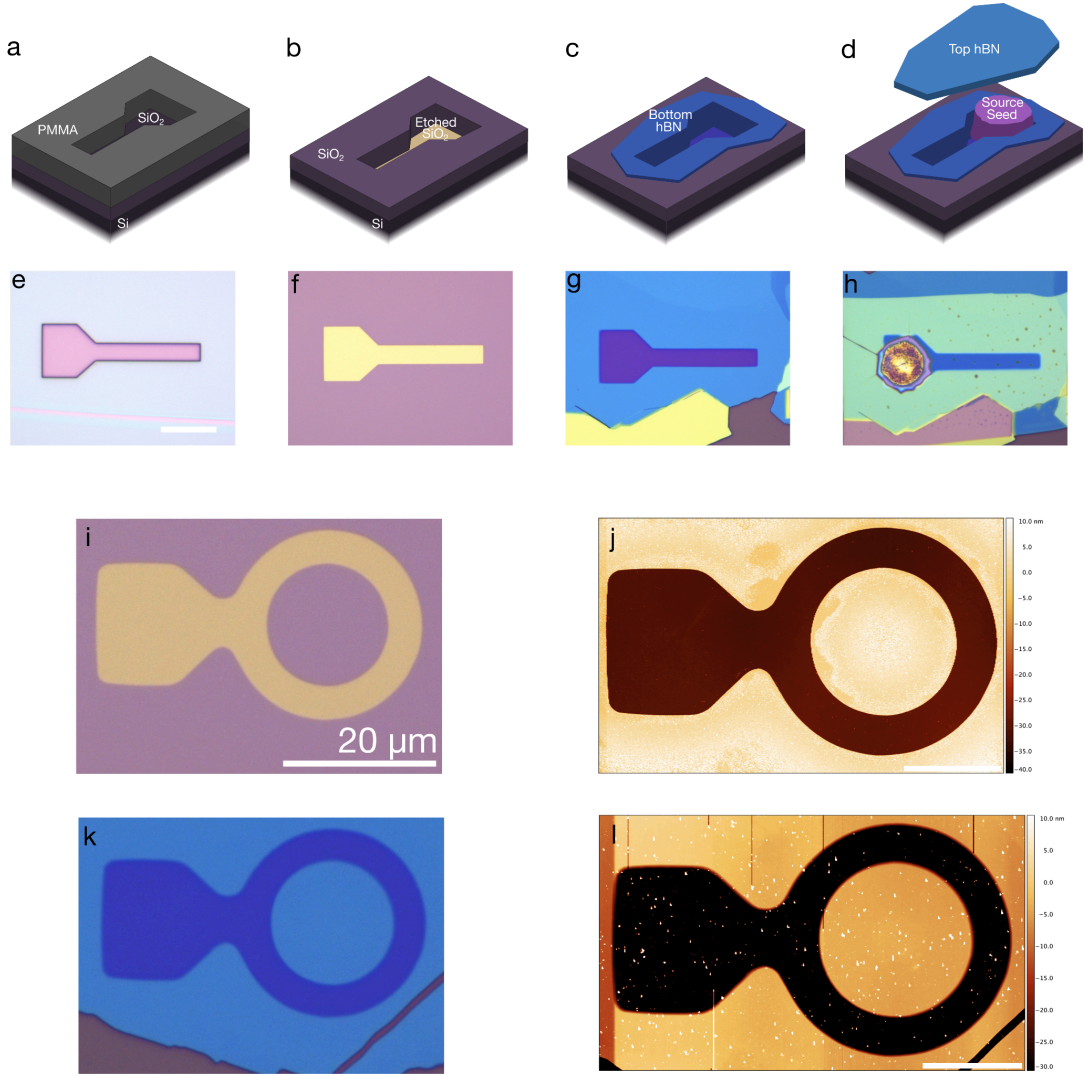

**Figure S1.** Depiction of van der Waals injection mold structure fabrication process: **(a,e)** PMMA mask creation, **(b,f)** etching and mask removal, **(c,g)** transfer bottom hBN, **(d,h)** transfer source flake and top hBN. **(i,j)** show optical and AFM topography image of $SiO_2$ etched trench; **(k,l)** show optical and AFM topography image of hBN transferred into trench.

**vdW Injection Molding Process**

Once the final vdW injection mold is fabricated, the injection molding process follows as similar to described in reference[3] where a top substrate (VWR coverglass or University Wafer Sapphire) is laid on top of the injection mold $SiO_2$/Si chip and fixed relative to the chip using kapton tape. Pressure is applied by a sapphire hemisphere driven by a translation stage.

During the molding step we bring the top pressing substrate into contact with the bottom substrate, leading to thin film interference color changes. We increase pressure until we achieve a uniform color in a range between magenta (lowest pressure) and yellow (highest pressure) (see SI video 2). Color differences correspond to differences in pressure and distance between the top and bottom substrates. We estimate that we are able to apply pressures within 10-100 MPa[3].

Aside from the samples shown in the main text, we have made additional samples in more complicated shapes (Fig. S2) including nanowires (Fig. S3).

The geometry of the molded crystal matches the geometry of the mold within the resolution of our optical microscope (Fig. S4).

This approach can be insensitive to the starting volume of material, provided it is larger than the trench volume. Excess material tends to be squeezed out of the hBN layers upon increasing pressure as shown in SI video 2, while still trapping material in the region area of the etched $SiO_2$ trench. However, sometimes, excess material can remain. As seen in Fig. 2(b) and Fig. S2, excess material outside of the etched $SiO_2$ region remains near the position of the source. We speculate that in these cases, the thickness of the original oxide prevents the applied pressure from squeezing out excess material.

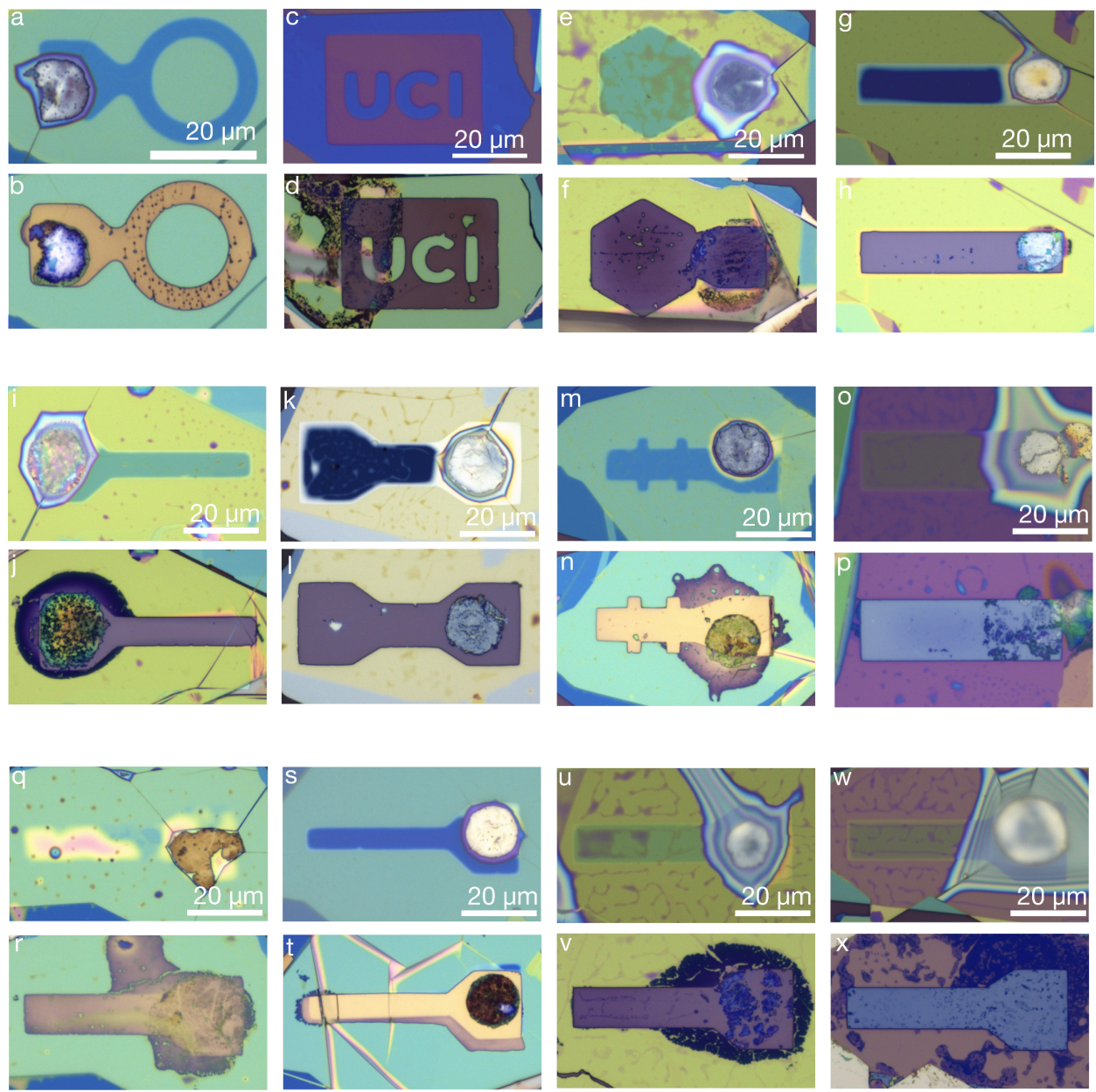

**Figure S2.** Bismuth samples not shown in the main text. Each of the 3 rows has an upper and lower optical image. The upper (lower) image is the completed vdW injection mold before (after) the vdW injection molding process.

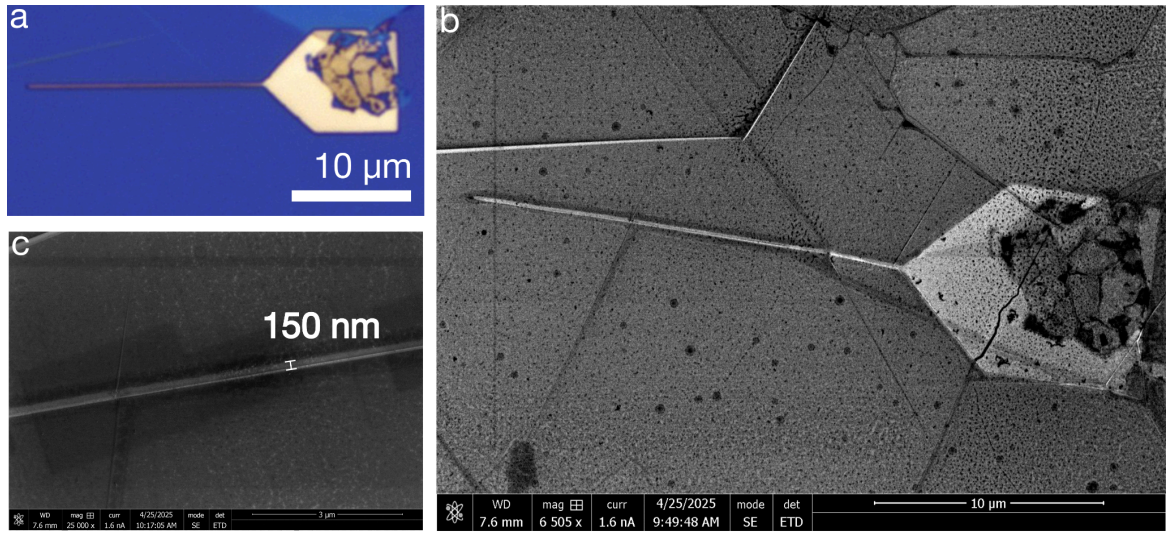

**Figure S3.** SEM images of bismuth nanowire LV18 showing a width of 150 nm.

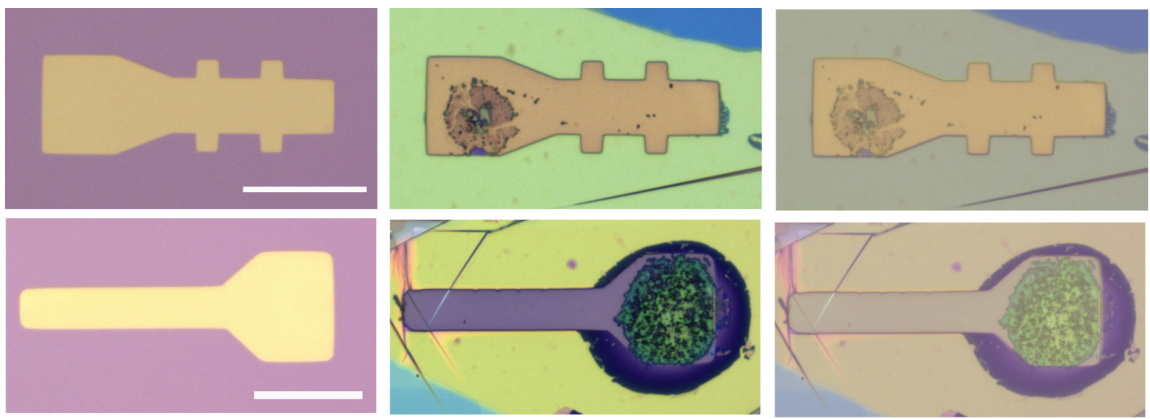

**Figure S4.** Trench Lateral Area versus sample lateral area. Each row consists of optical images of: etched SiO$_2$ trenches (first column), sample fabricated using that trench (second column), and image of the two superimposed on top of each other (final column). The areas agree within the optical resolution provided by our microscope. Scale bar is 20 µm.

**Sn and In Samples**

The fabrication of tin and indium samples follows a similar process to bismuth. For tin, the vdW injection mold is first heated to 220°C, 12°C below the melting point. Finally, the injection mold is allowed to cool while maintaining pressure. For indium, source flakes are picked up with a PC stamp with slits cut into the PC around the PDMS block. This allows for the source flake to be put down at a lower temperature of ~150 °C, under the melting point of indium. The encapsulated indium mold is then heated to the melting point of indium 156 °C before applying pressure during the injection molding process.

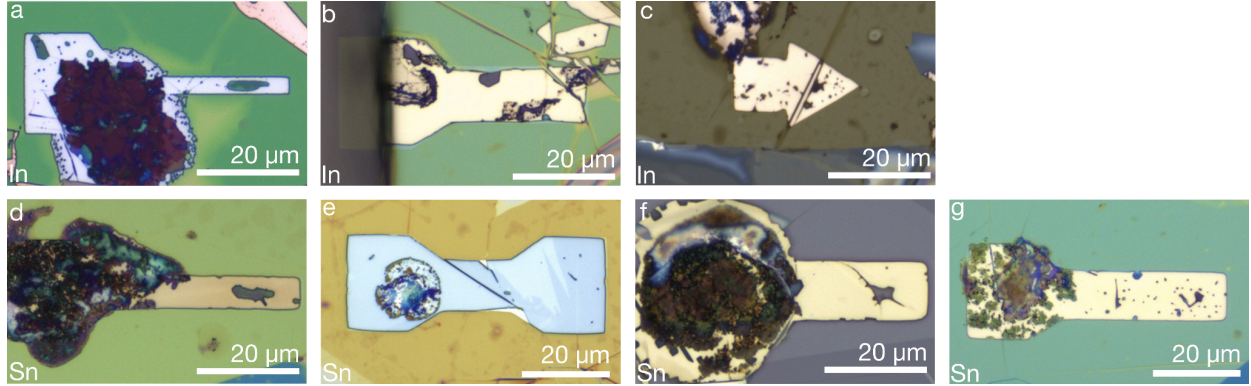

**Figure S5**. Optical images of van der Waals injection molded indium (a-c) and tin (d-g) samples. Faceted edges are observed along the sides and within internal voids. These features grow with time and when heating.

**Electronic transport device fabrication process.**
After fabricating the vdW injection molded crystals, we remove the crystal from the etched $SiO_2$ trench using polycarbonate/PDMS transfer techniques[2] and transfer it onto an unetched $SiO_2$/Si chip (Fig S5 a-c). Next, PMMA (Kiyaku PMMA 950 A5, 500 nm, 2000 RPM, 2 minutes) is spun onto the sample and electrodes are patterned using electron beam lithography (Fig S5 c-e). Then, the exposed regions of the top hBN are etched away using SF6 reactive ion etching (Trion ICP and RIE). After SF6 etching, the crystal is transferred into an electron beam evaporation system with a built in ion mill. The oxide layer on the exposed bismuth contacts is ion-milled away immediately before a thin layer of metal (5nm Ti, 15nm Au) is deposited. A thicker layer of metal (5nm Cr, 95nm Au) is then later deposited in a separate system (Angstrom EvoVac) (Fig S5(f-g)). PMMA and excess gold are then removed in a lift off procedure.

Transport measurements are done in a Cryomagnetics CMAG 12T closed-cycle cryostat using standard low-frequency lock-in (Stanford Research Systems SR830) techniques.

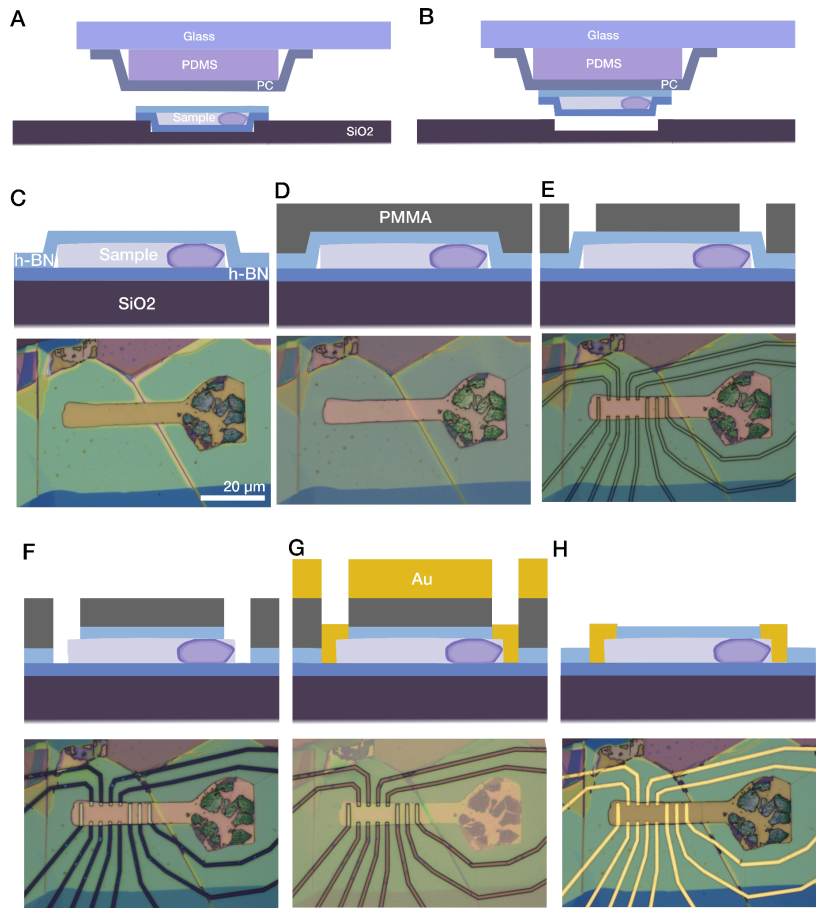

**Fig S6.** Electronic transport device fabrication process. (a-c) PC/PDMS transfer process to remove sample from etched trench onto SiO$_2$/Si substrate; (c-e) electron beam lithography to define electrodes; (f-g) reactive ion etching, ion milling, and metal evaporation of contacts; (h) lift off

**Height Analysis**

We determine the thickness of our samples from AFM (Park NX10) topography images. After injection molding, we remove the samples from the etched trench and transfer them onto unetched SiO$_2$/Si wafers (see Fig S5a-c) before imaging. Thickness is measured as the difference in height from the hBN/Sample/hBN regions to the hBN/hBN only regions.

We level the hBN/hBN background and set it as a zero reference point. We then mask off the hBN/hBN only region to collect height data from only the hBN/sample/hBN region (or region that we would make a device from). We also mask off impurities such as dirt on the outer surface of the hBN from the PC transfer process. We then fit a Gaussian to the resulting height histogram and extract the mean (average sample thickness) and standard deviation from this fit. The results for our bismuth samples are shown in SI Table 1.

| Sample | Mean Thickness AFM Height Distribution (nm) | Standard Deviation AFM Height Distribution (nm) | Percent Variation | Sample Area Considered (micron$^2$) | Orientation | Top substrate |
|---|---|---|---|---|---|---|
| S29 | 83.17 | 2.67 | 3.21% | 520 | | |
| S30 | 64.54 | 6.26 | 9.70% | 627.2 | 110 | |
| S31 | 80.585 | 2.92 | 3.62% | 782.31 | | |
| S42 | 64.78 | 1.425 | 2.20% | 220 | | |
| S44 | 61.28 | 2.11 | 3.44% | 273 | | |
| S45 | 58.877 | 2.62 | 4.45% | 209 | 111 | |
| S47 | 35.39 | 2.59 | 7.32% | 300 | | |
| S51 | 105.27 | 0.63 | 0.59% | 48 | | |
| S54 | 51.95 | 1.875 | 3.60% | | | Sapphire |
| S55 | 45.87 | 1.13 | 2.46% | 60 | | Sapphire |
| S56 | 42.01 | 1.11 | 2.64% | 140 | | Sapphire |
| S57 | 48.73 | 1.24 | 2.54% | 105 | | |
| S58 | 54.51 | 0.737 | 1.35% | 112 | 110 | |
| S61 | 19.04 | 1.74 | 9.14% | 106.2 | | |
| S62 | 21.11 | 1.09 | 5.16% | 123.5 | 110 | |
| S63 | 24.967 | 0.84 | 3.36% | 130 | 111 | |
| S65 | 36.78 | 0.615 | 1.67% | 110 | | Sapphire |
| S67 Upper Half | 11.51 | 0.688 | 5.98% | 80 | | |
| S67 Bottom Half | 11.02 | 2.02 | 18.3% | 128 | | |
| S68 | 31.18 | 1.83 | 5.87% | 189 | 111 | Sapphire |
| S68 12um | 31.16 | 1.34 | 4.3% | 126 | | |
| S69 | 37.15 | 1.92 | 5.17% | 184.85 | | |
| S70 | 29.69 | 1.95 | 6.57% | 275.67 | | |
| S71 | 28.88 | 2.97 | 10.28% | 644 | 110 | |
| S76 | 37.906 | 0.90 | 2.37% | | | Sapphire |
| S79 | 29.48 | 3.11 | 10.55% | 153 | 111 | |

*SI Table 1.* Table containing information on sample thickness (mean, standard deviation, and percent variation), area considered for the scan, orientation and top squeezing substrate used. Sample orientation determined through electron backscatter diffraction (EBSD) and transmission electron microscope (TEM) selected area diffraction measurements.

**Height Variation Across Width and Length Analysis**

To analyze the height variation across the width of our samples, we calculate the percent difference between the edges of the sample and the center of the sample. To calculate the edge and center values, we take height distributions localized to the edge or center regions and from that extract the mean values. We then also get an average percent difference of 1.65% and an average slope of 0.18 ± 0.15 nm/μm.

| Sample | Average Edge 1 Thickness (nm) | Average Center Thickness (nm) | Average Edge 2 Thickness (nm) | Average Percent Difference | Center to Edge Difference (um) | Average Slope (nm/um) |
|---|---|---|---|---|---|---|
| S51 | 106.5 ± 0.494 | 105.3 ± 0.434 | 105.3 ± 0.49 | 0.57% | 3 | -0.2 |
| S54 | 52.88 ± 1.355 | 51.19 ± 1.871 | 52.42 ± 1.492 | 2.81% | 3.4 | -0.42 |
| S63 | 25.39 ± 0.787 | 26.68 ± 0.546 | 26.6 ± 1.141 | 2.63% | 5 | 0.137 |
| S65 | 36.35 ± 0.64 | 36.85 ± 0.5 | 36.79 ± 0.473 | 0.76% | 5 | 0.056 |
| S76 | 37.39 ± 0.872 | 37.95 ± 0.84 | 37.31 ± 0.79 | 1.46% | 4-7 | 0.109 |

*SI Table 2. Height variation along width for five samples. From this we calculate an average percent difference of 1.65% and an average slope (measured from edge to center) of -0.064 nm ± 0.24 nm. Considering absolute slope, we get 0.18 ± 0.14 nm.*

| Sample | Average Edge 1 Thickness (nm) | Average Center Thickness (nm) | Average Edge 2 Thickness (nm) | Average Percent Difference | Center to Edge Difference (um) | Average Slope (nm/um) |
|---|---|---|---|---|---|---|
| S51 | 105.7 ± 0.52 | 105.4 ± 0.65 | 106.1 ± 0.89 | 0.47% | 4.25 | -0.1176 |
| S54 | 51.57 ± 1.46 | 49.07 ± 1.55 | 51.83 ± 1.32 | 5.15% | 8.8 | -0.284 |
| S63 | 26.46 ± 0.81 | 26.02 ± 0.70 | 26.74 ± 0.88 | 2.20% | 5.89 | -0.0985 |
| S65 | 37.41 ± 0.42 | 36.71 ± 0.48 | 36.46 ± 0.57 | 1.29% | 5.5 | -0.0409 |
| S76 | 37.54 ± 0.86 | 38.19 ± 0.71 | 38.29 ± 0.94 | 1.24% | 4-7 | 0.1724 |

*SI Table 3. Height variation along length for the same five samples. From this we calculate an average percent difference of 2.08% and an average slope (measured from edge to center) of -0.075 nm ± 0.17 nm. Considering absolute slope, we instead get 0.14 nm ± 0.17 nm.*

For S76, which has a hall bar geometry, we AFM different regions denoted as center, top, and hall bar tab 1 separately and average based weighted by area.

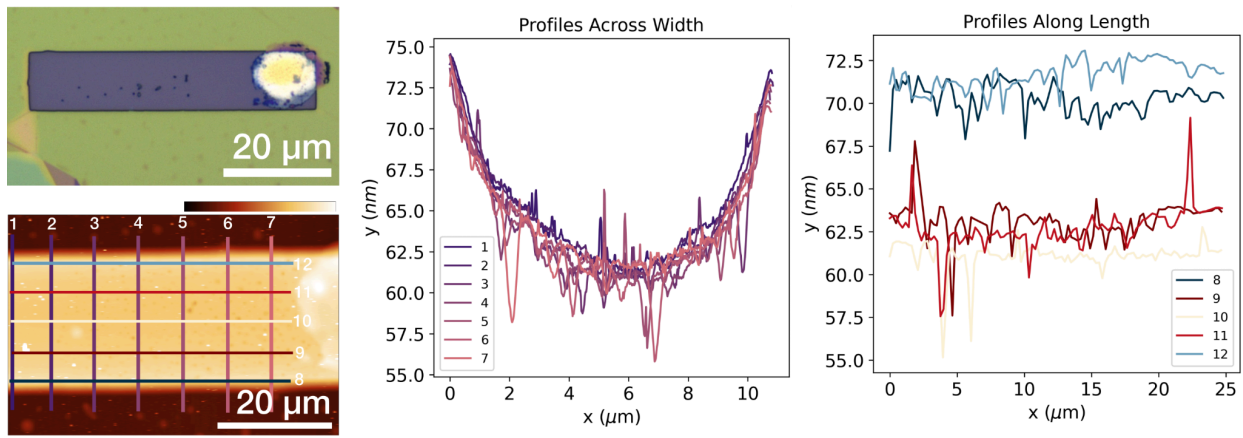

**Figure S7**. Height variation across width versus length. Here are plotted line cuts along the width of the sample (1-7) and line cuts along the length of the sample (8-12). Line cuts 1-7 show a concave height profile with a difference of about 14nm from edge to center. Line cuts 8-12 instead show a flatter height profile.

**Surface Characterization**

    To characterize the surface roughness of the molded crystals, we mechanically remove the top hBN using a polyvinyl chloride/PDMS stamp[3–5] and AFM the surface directly. Within a single terrace (an area around 1 µm$^2$), we measure the root mean square surface roughness (Gwyddion). For a tin sample (Sn02), we find values of 0.094 to 0.11 nm (Fig S7 a-b, d-e). For a bismuth sample (S55), we find values of 0.11 nm (Fig S7 c,f) agreeing with previous results[3]. As a comparison, the surface roughness of hBN is 0.05 nm.

    We also get a clearer image of the terrace step height. In Fig S8 we measure the terrace step height on a bismuth sample (S58) at 31 different locations to get an average step height of 0.33 nm ± 0.05 nm which is consistent with the 110 orientation. By comparison the 111 orientation has a step height closer to 0.39 nm[6]. In this manner, AFM topography of terrace height is another tool we use to determine crystallographic orientation.

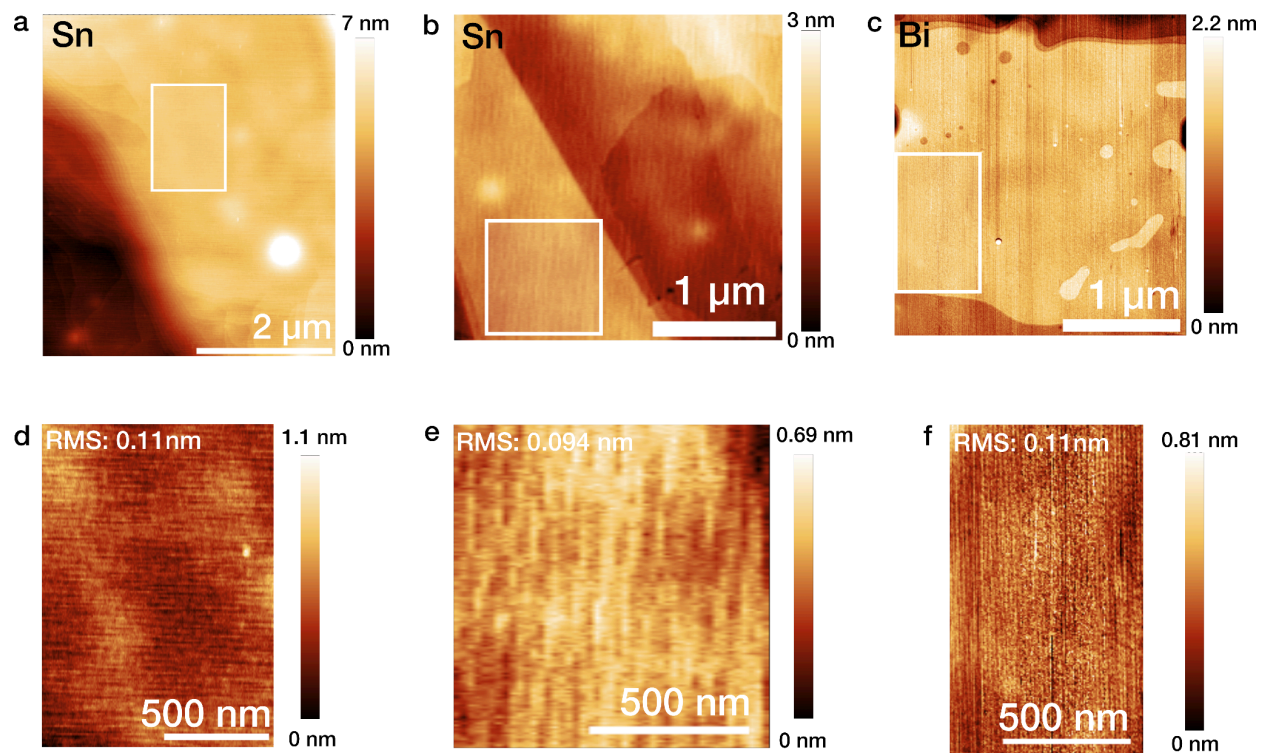

**Figure S8.** AFM of vdW molded crystal surface to determine surface roughness. (a-c) show AFM scan of larger region and (d-f) show reduced area in a single terrace used for surface roughness analysis.

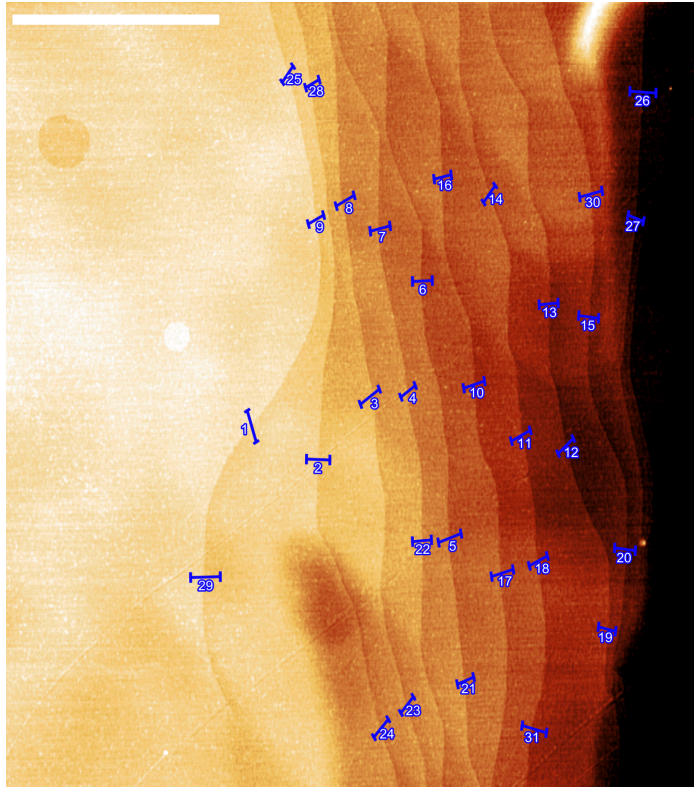

**Figure S9.** Line cuts across different terraces to calculate an average step height of 0.33 nm ± 0.05 nm for a bismuth sample (S58) in main text Fig. 3f consistent with the terrace step height of (110) orientation[6]. Scale bar is 1 µm long.

**Crystallinity**

We characterize the crystal domain size and crystallographic orientation using electron backscatter diffraction (EBSD) and transmission electron microscope (TEM) selected area diffraction measurements.

To prepare samples for TEM, we transfer the sample out of the trench and onto a TEM grid (silicon nitride with 1 um holes) using polycarbonate/PDMS transfer techniques[2]. For two samples (S45 and S68) we removed the top hBN before transferring so that the resulting structure is Bi/hBN structure. We kept two samples (S79 and S65) encapsulated, studying a hBN/Bi/hBN structure. The resulting diffraction patterns are shown in SI Fig S9 and main text Fig 3e showing sharp peaks suggesting high crystallinity. The diffraction patterns shown are consistent with the bismuth 111 orientation. Note, that additional hexagonal diffraction patterns are observed from the hBN layers.

For EBSD, we perform diffraction measurements on five encapsulated and top-hBN removed samples (Tescan GAIA3 SEM-FIB). IPF-Z and IPF-Y maps of bismuth orientation are plotted using MTEX.

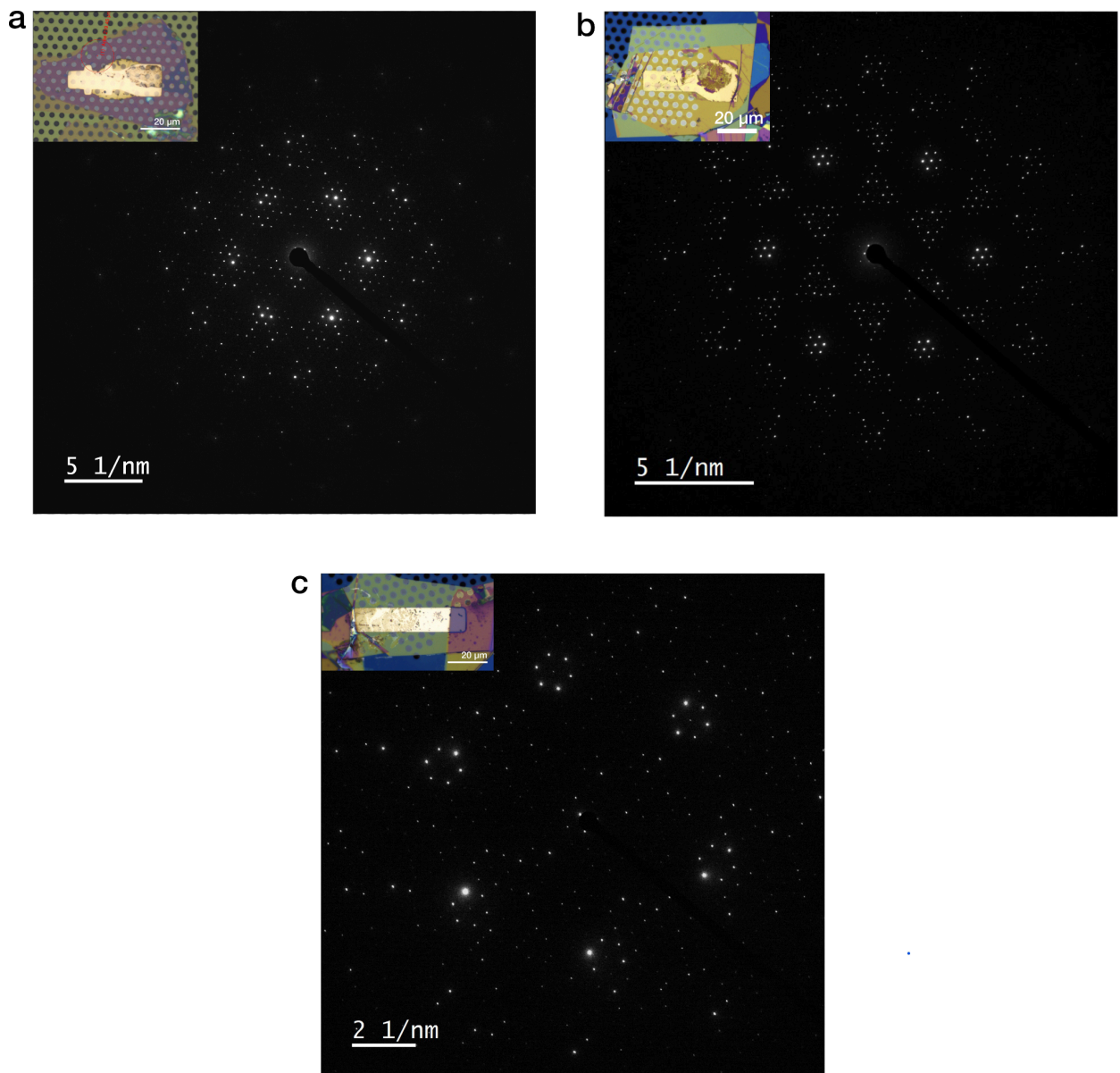

**Figure S10**. Selected area diffraction images of additional bismuth samples. **(a,b)** Diffraction on hBN/Bi/hBN samples show sharp diffraction peaks as well, indicating high crystallinity in that region, though the subsequent pattern is more complex (S79 in a and S65 in b). **(c)** Diffraction on a Bi/hBN sample (S45).

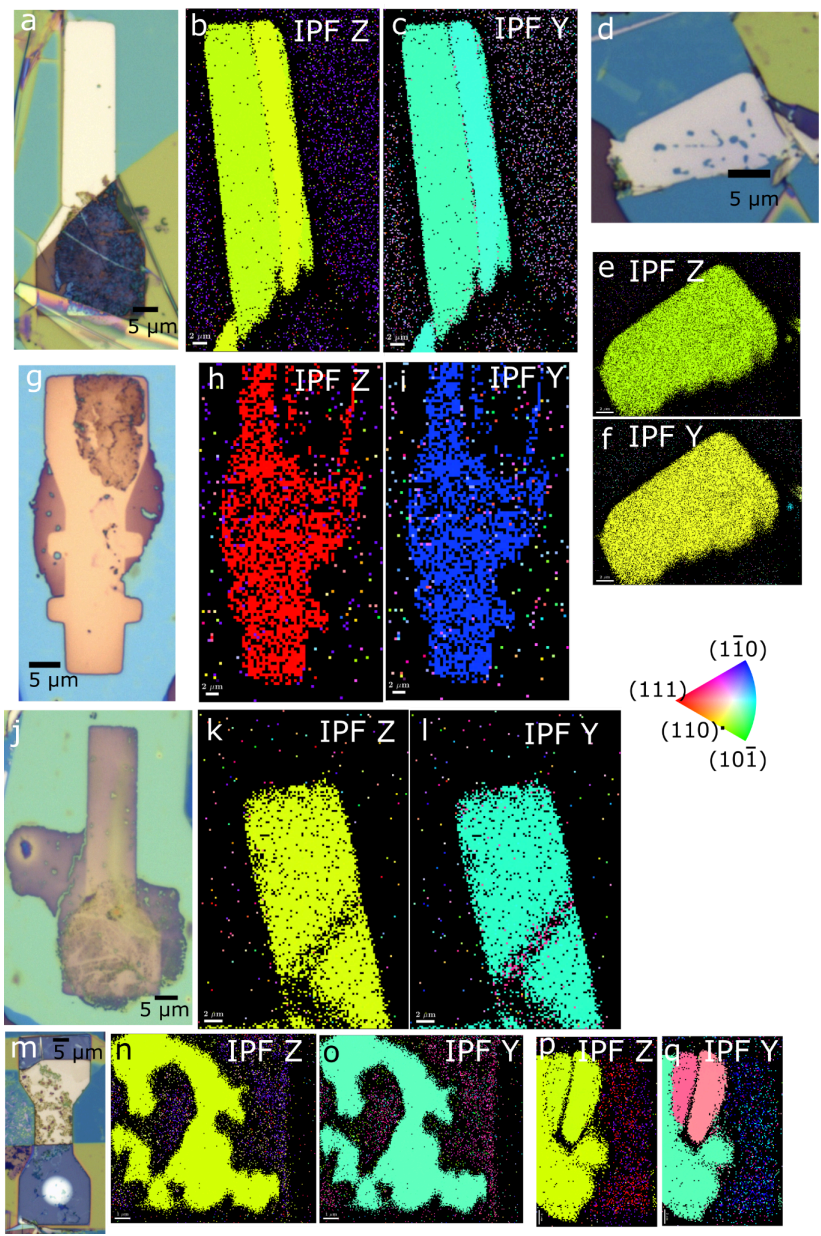

**Figure S11.** EBSD of five different vdW injection molded bismuth samples. IPF-Z and IPF-Y maps of bismuth orientation are plotted using MTEX. a-f) EBSD of samples with the top hBN removed (S58 in a-c and S71 in d-f). (g-l) EBSD of hBN encapsulated samples (S79 in g-i and S62 in j-l). i-l) EBSD of a sample with top hBN removed (S30) at different locations. The dark regions in n) and o) are in regions where remnants of an oxide shell from the source flake are on top of the bismuth, weakening the signal in those regions.

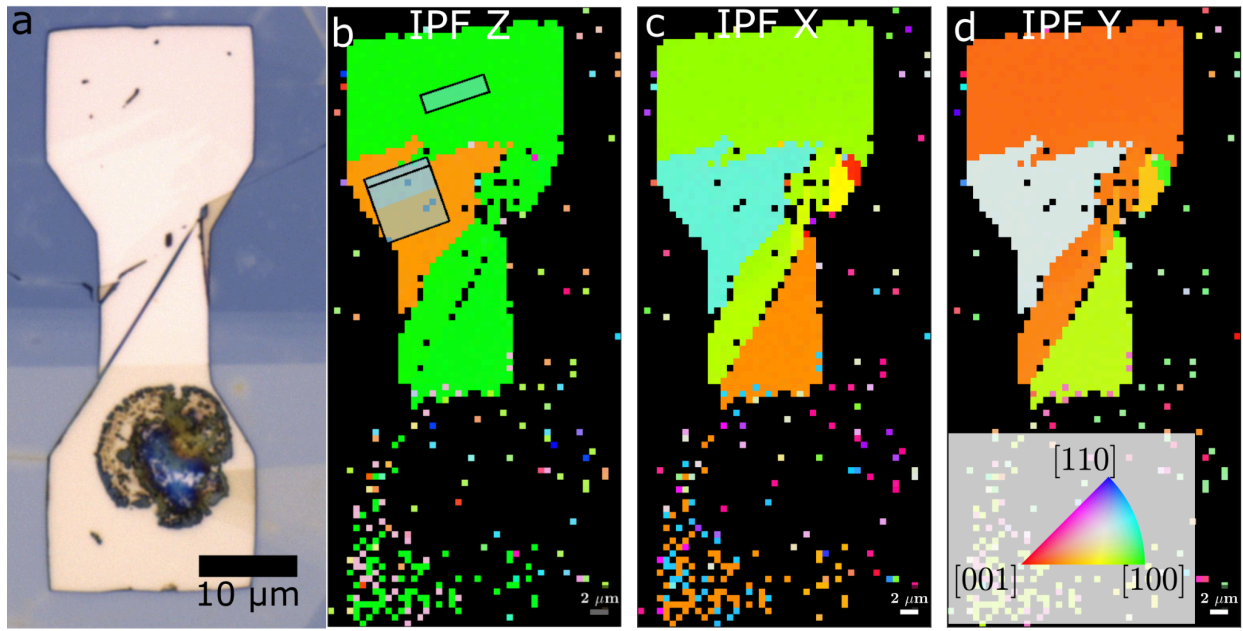

**Figure S11.** a) Optical image of hBN and trench molded white Sn with the left behind blue oxide shell. b) EBSD of IPF Z map shows □-Sn sample with 2 out of plane rotations. Light blue rectangles visualize the lattice plane facing the z-axis. c) & d) IPF X and IPF Y maps show the larger grain contains some in-plane rotations. The diffraction signal near the oxide shell is weakened in all of the EBSD maps.

**Hermetic Sealing**

As a test of the effectiveness of the hBN/hBN interface at providing environmental protection during the injection molding process, we take a bismuth injection molded crystal (S88) and melt it in ambient conditions without applied pressure. We observe the crystal coalesce into three clusters with thicknesses 150-350 nm, which is consistent with a lack of a stabilizing oxide shell and bismuth's high surface tension[7] (Fig S12c,e). The regions where the injection molded crystal were formerly located are also optically similar to the nearby crystal-free hBN/hBN region.

AFM of regions where the vdW molded crystal was formerly located are topographically similar to the nearby hBN/hBN regions that never contained any injection molded material (Fig S12 d-e). One possible interpretation is that the molded bismuth did not form a sufficiently thick oxide (or any oxide) layer that would resist the movement of the molten material in a similar fashion to the oxide shell that initially surrounds our flakes of source material. For comparison, a ~7 nm thick remnant of the oxide shell that initially surrounded the source material is visible in Fig S12d.

In summary, we observed crystal coalescence under remelting and similarity between regions that formerly held material and regions that did not through optical images and AFM. These observations are consistent with the molded bismuth having not oxidized or having minimally oxidized due to the hBN/hBN interface forming an effective hermetic seal.

**Clean Interfaces**

Injection molding also tends to clean trapped organic residue present from the vdW stacking process. Observable in Fig S12d is the outline of the mold region where the source flake was placed. The interior region is free of bubbles that by contrast populate the adjacent hBN region outside of the mold. An explanation is that the injected molten material picks up and carries away residue that it comes into contact with.

We also observe features that form an outline of the ring-mold geometry. Also present are features that trace a concentric ring in the middle of this outline (Fig S12e-f). We speculate that these features are polymer residue from the transfer process that were pushed to the edge of the ring-shaped trench during the injection molding or trapped in the unfilled areas of the crystal (Fig S12b).

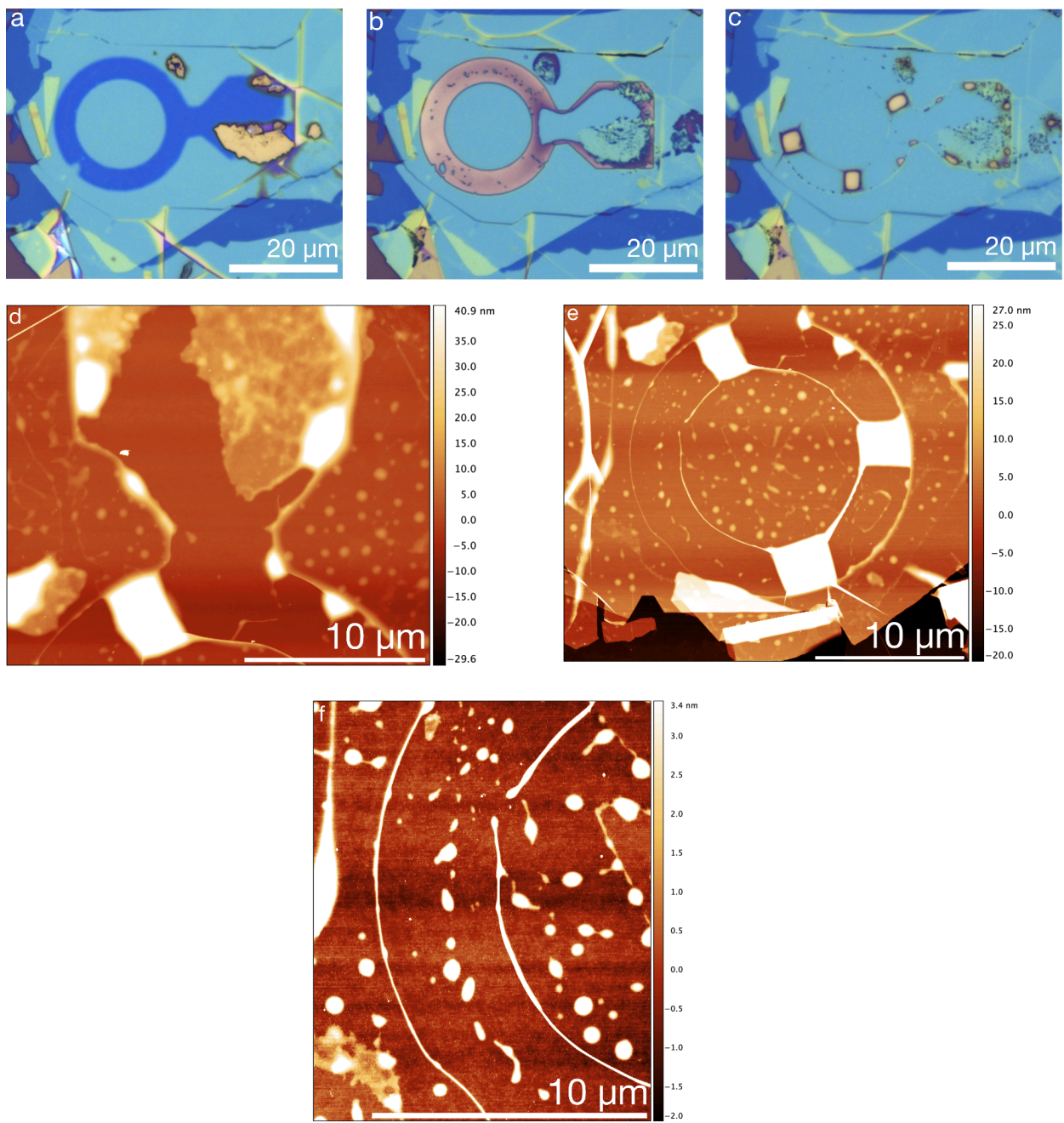

**SI Figure 12.** (a) Optical image of vdW injection mold structure before injection molding. (b) hBN encapsulated injection molded crystal (S88) transferred onto unetched SiO$_2$ wafer. Optically visible on the right are remnants of the oxide shell that initially surrounded the source material flake. (c) Result after melting S88 without pressure applied. (d-f) AFM topography images of the hBN encapsulated surface post melting.